\documentstyle[epsfig]{elsart}
\def\be{\begin{equation}}
\def\ee{\end{equation}}
\begin{document}
\begin{frontmatter}
\title{Total Widths And Slopes From Complex Regge Trajectories}
\author[Pg]{S. Filipponi\thanksref{emsi}}, 
\author[Lnf]{G. Pancheri\thanksref{emp}} and 
\author[Pg,Bo]{Y. Srivastava\thanksref{emsr}}
\address[Pg]{Physics Department-University of Perugia and INFN-Perugia\\
via Pascoli, I-06100 Perugia, Italy.}
\address[Lnf]{Laboratori Nazionali di Frascati dell' INFN\\
 Frascati, Rome, Italy.}
\address[Bo]{Physics Department-Northeastern University\\
 Boston, Mass, USA.}
\thanks[emsi]{Silvana@pg.infn.it}
\thanks[emp]{Pancheri@lnf.infn.it}
\thanks[emsr]{Srivastava@pg.infn.it}
\begin{abstract}
Maximally complex Regge trajectories are introduced 
for which both Re $\alpha(s)$ and Im $\alpha(s)$ grow as 
$s^{1-\epsilon}$ ($\epsilon$ small and positive).
Our expression reduces to the standard real linear form as the imaginary part 
(proportional to $\epsilon$) goes to zero. A scaling formula for the total 
widths emerges: $\Gamma_{TOT}/M\rightarrow$ constant for large M, in very good 
agreement with data for mesons and baryons. The unitarity corrections 
also enhance the space-like slopes from their time-like values, thereby 
resolving an old problem with the $\rho$ trajectory in $\pi N$ charge 
exchange. Finally, the unitarily enhanced intercept,  
$\alpha_{\rho}\approx 0.525$, \nolinebreak  is in good accord with the 
Donnachie-Landshoff total cross section analysis. 
\end{abstract}
\end{frontmatter}
\section{Introduction}
It is well known that hadronic mass spectra can be described quite reasonably 
through real, linearly rising Regge trajectories. Veneziano model and hadronic 
string models are anchored on such a hypothesis.
On the other hand, unitarity and analyticity require
the addition of an imaginary part thereby providing total widths for the 
resonances as well as giving some curvature to the trajectories. The modest 
aim of our work is as follows. Using the positivity of the imaginary part 
and the 
positivity of the slope, it is observed that the real parts of trajectories can 
rise no faster than linearly. We then introduce the  notion  of maximally 
complex trajectories for which both the real and the imaginary parts grow    
in the same way, i.e.,
\be
\left . 
\begin{array}{l}
\mbox{Re}\: \alpha(s)\\
\mbox{Im}\: \alpha(s)
\end{array} \right \} 
 \longrightarrow\: s^{1-\epsilon}
\left\{  \begin{array}{l}
\cos \pi \epsilon \\
\sin \pi \epsilon 
\end{array} \right .
\hskip8mm \mbox{for}\;s\rightarrow \infty,  
\label{hypo}
\ee
where $\epsilon$ is small but positive. Physically, our hypothesis embodies
the dictum that {\it ``strong interactions are as strong as they can be''}.
It is also satisfactory mathematically since a small imaginary part can only 
curve $|\alpha(s)|$ downwards from linearity. The limit $\epsilon$ equals zero 
corresponds to no imaginary part and a linear real trajectory.
Implications of the resulting deviations from linearity are studied for
the real parts to discover significant changes in the intercept, in 
agreement with experimental observations. On the other hand, analysis of
the imaginary parts leads us to a pleasing result that total widths of high 
mass resonances grow linearly with mass, a fact which we show to be true
experimentally, \cite{DP96}, both for mesons and for baryons.
\section{ Formalism}   
Under the assumption that the complex pole function $\alpha(s)$ has: \par
\begin{enumerate}
\item 
\begin{itemize} 
\item  a single zero at $s\ =\ -s_0$
\item with a positive derivative;
\end{itemize}
\item a unitarity branch cut  for $s\ge\Lambda^2$, 
with $\mbox{Im}\: \alpha (s)\ge 0$;
\end{enumerate}
it is possible to show that its asymptotic behavior 
is bounded as follows
\be
\left |\frac{1}{s}\right |\: \leq \:
\left |\alpha (s) \right |\: \leq \:|s|
\hskip8mm \mbox{for}\; |s|\rightarrow \infty
\label{bound}
\ee
\indent To show that  
$|\alpha(s)|\leq |s|$, let us assume the contrary, i.e., 
that $|\alpha(s)| \rightarrow |s|^{1+\delta}$,  
($\delta >0$), as $|s|\rightarrow\infty$. Now we write 
a dispersion relation for the inverse function $1/\alpha(s)$, 
which  has a single pole at $s=-s_0$
(with a positive residue) 
\be
\frac{1}{\alpha(s)}=\frac{A}{s+s_0} - \int _{\Lambda ^2}^{\infty} 
\frac{ds'}{\pi} \frac{\mbox{Im}\: \alpha (s')}{(s'-s-i 0^+)
|\alpha(s')|^2}
\label{drinv}
\ee 
Taking the asymptotic limit of the previous Eq.(\ref{drinv}), one has 
\be 
\frac{1}{|\alpha(s)|} \longrightarrow \frac{1}{|s|}\left [ A + 
\int _{\Lambda ^2}^{\infty} 
\frac{ds'}{\pi} \frac{ \mbox{Im}\: \alpha (s')}{|\alpha(s')|^2} \right ]
\label{drinvas}
\ee
Since, $A$ is positive and 
the integral in Eq.(\ref{drinvas}) is positive and convergent, 
$\frac{ \mbox{Im}\: \alpha }{|\alpha| ^2}<\frac{1}{s^{1+\delta}}$, 
\nolinebreak the sum
can not vanish. Hence, we have proof ad absurdum. A similar 
derivation holds also for the 
lower bound condition: $|\alpha(s)|\ge |1/s|^{1+\delta}$\par
Our hypothesis of maximum complexity may be written in the following
suggestive ``renormalized'' form
\be
\alpha(s)\ \rightarrow\ {\bar \alpha}(s)\ Z(s;\epsilon)
\label{maxco1}
\ee 
where, ${\bar \alpha}(s)$ symbolizes the real, linear ``unrenormalized'' 
trajectory and $Z(s;\epsilon)$ is the complex``renormalization'' factor
with the boundary conditions
\be
\begin{array}{l}
Z(s;\epsilon=0)=1 \\
|Z(s;\epsilon)|\rightarrow|s|^{-\epsilon}
\hskip5mm \mbox{as}\;|s|\rightarrow\infty
\end{array}
\label{maxcol2}
\ee
The above Eq.(6) makes both Re and Im $\alpha(s)$ to grow like 
$s^{1-\epsilon}$, and the 
asymptotic phase of $\alpha$ goes to a constant. This leads total widths 
to grow like $\sqrt s$ as shown below.\par
Writing the contribution of a trajectory to an angular momentum $J$ 
partial wave amplitude as
\be
T(J,s)=\frac{\beta(s)}{J - \alpha(s)},
\label{amp1}
\ee 
and comparing it to a Breit-Wigner form for a given $J$ resonance, it
is easily seen that for a resonance of mass $M$ and total width 
$\Gamma_{TOT}(M)$
\be
\mbox{Im}\:\alpha(M^2)\;\;\approx\;\; \mbox{Re}\:\alpha'(M^2)
\:\: M \:\:\Gamma_{TOT}(M)
\label{amp2}
\ee
Since for small $\epsilon$, Im $\alpha(s)$ is proportional to $\epsilon$
and grows for large $s$ as $s^{1-\epsilon}$, we have the approximate result
\be
\frac{\Gamma_{TOT}(M)}{M}\: \longrightarrow\:  \tan(\pi \epsilon)\hskip8mm  
\mbox{for}\; 
 \mbox{large}\; M
\label{amp3} 
\ee
\section{A Model for Maximally Complex Regge Trajectories}
Let us write a dispersion relation for  $\eta (s)=
\log \frac{\alpha (s)}{s+s_0}$, normalize $\alpha (s)$ at a 
threshold called $\Lambda$ and impose that it has a zero at $-s_0$:
\be
\alpha(s)=\left (\frac{s+s_0}{\Lambda^2+s_0}\right )\:\alpha(\Lambda^2)\; 
e^{\left [(s-\Lambda^2)\:\int_{\Lambda^2}^{\infty} \frac{ds'}{\pi} 
\frac{\phi(s')-\phi(\Lambda^2)}{(s'-s-i\delta)(s'-\Lambda^2)}
\right ]}
\label{dr}
\ee
where $\phi(s)=\mbox{Im}\:\eta (s)$.\par 
A very simple model consistent with positivity, the threshold behavior and a 
constant asymptotic limit for the phase is given by
\be
\phi(s)-\phi(\Lambda^2)=\pi \epsilon \Big(1-{\Lambda^2  \over s}\Big)^p 
\label{model}
\ee
with $\epsilon$ positive and $p=( \alpha(\Lambda^2)+1/2)$, as 
required by analyticity of the trajectory. 
Actually the phase $\phi(s)$ is zero at threshold. (In the above, for simplicity
we are considering equal masses for the two scattering particles).\par
For arbitrary $p$, the integral in Eq.(\ref{dr}) is rather messy and shall be treated
in detail elsewhere. For illustration, we shall explicitly
exhibit the  example of $p=1/2$, which is of interest
for an S- wave resonance near threshold. For this case, 
 the integrals in Eq.(\ref{dr}) can 
be done explicitly for the three different regions:
\begin{itemize}
\item {\bf Region I}: $s>\Lambda^2$. \par\noindent 
For $s>\Lambda^2$, $\eta_I$  is complex. Hence  
both the real and the imaginary part of the pole are different from zero. 
\be
\begin{array}{l} 
\mbox{Re}\: \alpha_I (s)\\
\mbox{Im}\: \alpha_I (s)
\end{array}
=\left [ \frac{s+s_0}{\Lambda^2+s_0} \right ] \:
\alpha(\Lambda^2)\; e^{\mbox{Re}\: \eta_I(s)}\times  
\begin{array}{l}  
\cos\:\mbox{Im}\:\eta_I(s) \\ 
\sin\:\mbox{Im}\:\eta_I(s)
\end{array} 
\label{t1}
\ee
where $\eta_I$ is given by
\be
\mbox{Re}\: \eta_I(s)=-\epsilon \sqrt{1-\frac{\Lambda^2}{s}}\;\:\ln \left[ 
\frac{1+\sqrt{1-\frac{\Lambda^2}{s}}}{1-\sqrt{1-\frac{\Lambda^2}{s}}}
\right ]
\label{f1}
\ee
and 
\be
\mbox{Im}\: \eta_I(s)=\pi \epsilon \sqrt{1-\frac{\Lambda^2}{s}}
\label{fi1}
\ee
\item {\bf Region II}: $0<s<\Lambda^2$. \par\noindent
In region II, the trajectory is entirely real since $\mbox{Im}\:\eta_{II}=0$.  
It is given by 
\be
\alpha_{II}(s)=\left [\frac{s+s_0}
{\Lambda^2+s_0} \right ] 
\:\alpha(\Lambda^2)\; e^{\mbox{Re}\: \eta_{II}(s)}
\label{t2}
\ee
where 
\be
\eta_{II}(s)=-2\epsilon \:\sqrt{\frac{\Lambda^2}{s}-1}\;\: \tan^{-1}
\frac{1}{\sqrt{\frac{\Lambda^2}{s}-1}}
\label{f2}
\ee
\item {\bf Region III}: $s<0$.\par\noindent
As in case II, the trajectory is entirely real with the 
real phase $\eta_{III}$ given by
\be
\eta_{III}(s)=-\epsilon \sqrt{1-\frac{\Lambda^2}{s}}\;\:\ln \left[ 
\frac{\sqrt{1-\frac{\Lambda^2}{s}}+1}{\sqrt{1-\frac{\Lambda^2}{s}}-1}
\right ] 
\label{f3}
\ee
\end{itemize}
Boundary conditions are satisfied and $\eta_{II}(0)=\eta_{III}(0)=-2\epsilon$.
\par
The model exhibits two main features. 
First, as our initial conditions required, the pole function 
has a positive imaginary part vanishing at threshold.
Secondly, the  slopes for the space and time-like regions 
do differ once the trajectories acquire an imaginary part. 
In fact,  as confirmed by experimental data in $\pi N$ charge exchange, 
the slopes are larger in the space-like region. 
\section{Imaginary Part}
Above threshold (region I for which $s>\Lambda^2$), the formula for the 
imaginary part may more conveniently be written as follows
\be
\mbox{Im}\: \alpha (s)=\tan \left ( \pi\epsilon\; 
\left [1-\frac{\Lambda^2}{s}\right]^p  \right )\; \mbox{Re}\:  \alpha (s)
\label{ima}
\ee
For high mass time-like limit ($\Lambda ^2/s\rightarrow 0$),  
Eq.(\ref{ima}) leads to a very simple expression
\be
\mbox{Im}\: \alpha (s)= \pi\epsilon \;\;  \mbox{Re}\: \alpha (s) 
\label{imaapp}
\ee
According to the previous Eq.(\ref{imaapp}), the imaginary 
part of the pole  would be essentially linear with the squared 
energy just as the real part. We verify this behavior for various
meson and baryon resonances. \par
In Figure 1, we analyze experimental data for the $\Delta$ resonances. 
On the left side we show the real part of the trajectory ($J$ vs $M^2$)
through resonance masses. On the right side we show 
the corresponding imaginary part (as given by Eq.(\ref{amp2})) 
employing the total widths of the $\Delta$ resonances,
versus the squared 
mass values of the $\Delta$ resonances. 
As we can see, the asymptotic limit  described in Eq.(\ref{imaapp}) 
holds for this system quite accurately. 
The slope of Re $\alpha$ is found to be around 
0.88 $GeV^{-2}$ and the slope of Im $\alpha$ is found 
to be around $0.17\: GeV^{-2}$ \nolinebreak confirming that the $\epsilon$ 
parameter in Eq.(\ref{imaapp}) is indeed quite small (about a few percent). 
\vskip1mm
\begin{figure}[thb]
\begin{center}
\epsfig{file=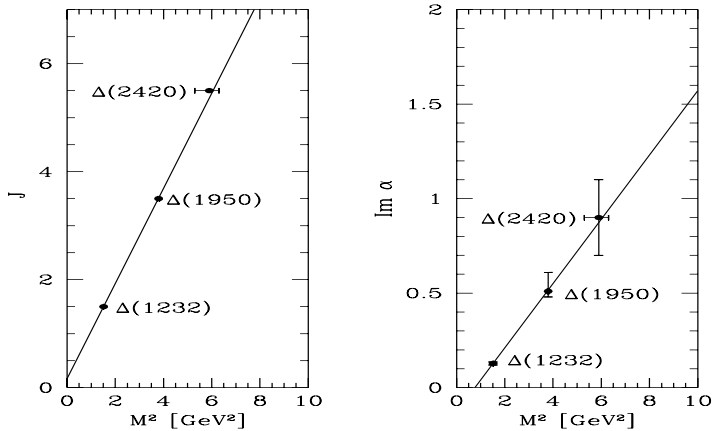,height=6cm,width=10cm}
\caption{The real [LEFT] and imaginary [RIGHT] part of the Regge trajectories 
for $\Delta$ resonances. The imaginary part of the pole is related to total 
widths  of the hadronic states through Eq.(\ref{amp2}).}
\end{center}
\end{figure}
Earlier T\"ornqvist, \cite{T91}, had found a similar systematics for the 
total widths of light mesons purely phenomenologically. Wherever our Regge 
analysis overlaps with his, we are in complete agreement. 
\subsection{Threshold  Correction Factors}
In order to be able to use all resonance data lying on a given
trajectory, including those
for which  threshold corrections are substantial, it is convenient to use the
following  expression for the total width, 
\be
\Gamma=\left ( \frac{J}{\alpha '\: M}\right ) \; \tan \left [ 
\pi\epsilon \;\rho(M,\Lambda_{\pm})\right ] 
\label{width}
\ee
where 
\be
\rho= \sqrt{\left[1-\frac{\Lambda^2_+}{M^2}\right]\;
\left[1-\frac{\Lambda^2_-}{M^2}\right]}
\label{rhoeff}
\ee
and $\Lambda_{\pm}$ $=$ $(m_1 \pm m_2)$. 
We note that 
\begin{itemize}
\item for a low mass resonance, such as the $\rho$,  the above expression 
differs 
from its asymptotic limit by a factor two;  
\item for the $\omega $ resonance the width is quite small since its primary 
decay 
via the $\rho \pi$ channel occurs only through the lower tail of the $\rho$. 
An effective mass for $\Lambda_+\ =\ 755$ MeV has been used; 
\item  For $\phi(1020)$, the effective threshold factor has been determined 
through
its P- wave decay ($85\%$) into $K\bar{K}$ and its S- wave decay ($15\%$) into
$\rho\ \pi$ channel;
\item Similarly, for $\Lambda(1520)$, the effective threshold has been 
determined 
through its decay $50\%$ each into $NK$ (P- wave) and $\Sigma\pi$ (S- wave).  
\end{itemize}
In Figure (2), we show a plot of $\Gamma_{TOT}/M\rho_{eff}$ for light
flavoured mesonic systems (the $\rho$, $\omega$, $\phi$ and $K^*$ recurrences)
and baryonic systems (the $N$, $\Delta$, $\Lambda$ and $\Sigma$ recurrences).
Since the input total widths ranged over a factor of a hundred, it is fair
to conclude that much order has been restored for each system and moreover
a not too dissimilar asymptotic value for all the light systems can be
inferred. 
\begin{figure}[bht]
\begin{center}
\epsfig{file=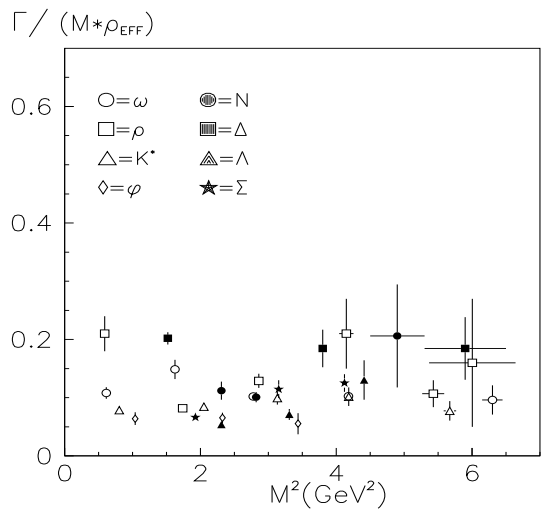,height=7cm,width=8.cm}
\caption{$\Gamma/(M\:\rho_{EFF})$ as a function of $M^2$ for light hadrons. 
$\Gamma$ is the total width, M the corresponding mass and 
$\rho$ is given by Eq.(\ref{rhoeff}). 
$\rho_{EFF}$, as discussed in the text, 
has been used for $\phi(1020)$ and $\Lambda(1520)$.}
\end{center}
\end{figure}
\section{Slopes for Space- and Time-like Regions}
As stated earlier,  slopes for the space and 
time-like regions of the real part of the trajectory differ 
once the trajectories acquire an imaginary part. The slope is seen to be larger 
in the space-like region, according to Eqs.(\ref{t1}),(\ref{f1}),(\ref{fi1}) 
and (\ref{f3}). \par
The most suitable example to check the  different values for the 
slope is given by the 
$\rho$ trajectory, where analyses of experimental data   
 have been performed in both regions. 
The time-like slope is found to be around $0.88-0.9 \: GeV^{-2}$, \cite{T90}. 
To infer the space-like slope, we turn to the high-energy limit of  
$\pi^-p\rightarrow\pi^0 n$. Here the $\rho$ trajectory 
 exchange ought to be a good approximation being the only one with  
the right t-channel quantum numbers. Earlier analyses of the differential 
cross section for the above scattering had revealed that the effective $\rho$ 
trajectory for negative t acquires a curvature and if linearity is imposed,
one is led to a larger value for the slope near s=0: $(0.97\pm 0.04)\:GeV^{-2}$ in 
 \cite{Bo74}, $(1.08\pm 0.03)\:GeV^{-2} $ in \cite{AC66} and 
$(1.00\pm 0.11)\:GeV^{-2}$ in \cite{Ho66}.
\vskip1mm
\begin{figure}[bht]
\begin{center}
\epsfig{file=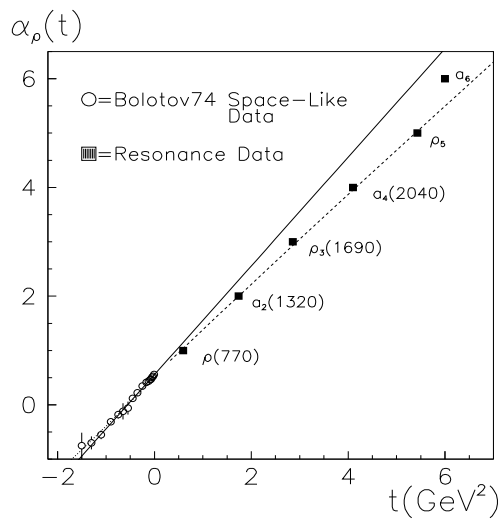,height=7cm,width=7.5cm}
\caption{$\rho$ trajectory in the space and time-like regions. Full 
($\epsilon=0$) and dashed ($\epsilon=0.03$) lines are 
predictions from  Eqs.(\ref{t1}),(\ref{f1}),(\ref{fi1}) and (\ref{f3}). A 
comparison with experimental data is given. For negative t values 
data are from the analysis in \cite{Bo74}.}
\end{center}
\end{figure}
Our model  describes the above picture quite well. 
 In Figure (3) experimental data for the masses of the $\rho$ 
resonances and the analyses in \cite{Bo74} for the space-like $\rho$ 
trajectory are plotted again with predictions from our model, 
Eqs.(\ref{t1}),(\ref{f1}),(\ref{fi1})
 and (\ref{f3}). 
The full line is for $\epsilon=0$, which would imply no variation 
for the slopes between the two regions. 
The dotted and dashed lines, which  are 
for $\epsilon=0.03$, describe data in both regions. 
 The threshold factor which appears in Eq.(\ref{t1}), was taken to be 
$\Lambda^2=4m^2_{\pi}$; and  $s_0=0.56\:GeV^{-2}$. \par
For the above parameters, the intercept $\alpha_{\rho}(0)$ is found to be 
0.525, which is in good accord with the effective intercept value 
$\alpha_{eff}(0)=0.547$ obtained through the total cross 
section analysis by Donnachie and Landshoff, \cite{DL}. 

\section{Conclusions}
In this work we have shown how constraints due to analyticity
and unitarity applied to a Regge pole function lead to rather
powerful predictions for the hadronic systems. An admittedly simple model 
for the phase consistent with positivity, the threshold and asymptotic
conditions  was presented which appears to work reasonably well.\par 
Since the imaginary part of the pole is directly related to the 
total widths of the hadronic resonance states, quantitative predictions 
can be made once we know the $\epsilon$ parameter, i.e. value of the 
asymptotic
phase.  For $\Delta$  and $\rho$ recurrences we found $\epsilon \approx 0.03$ 
to be in good agreement with data. Experimental data are consistent with 
our conjecture that the asymptotic phase has 
very nearly the same value for all of the light baryonic and mesonic 
sector. Moreover, whenever the threshold factor is 
close to one, the imaginary 
part of the trajectory appears to be approximately linear just as the real
part is. A more complete analysis shall be presented elsewhere \cite{fps98}.\par
Regarding the real part of the trajectory, we showed that the slopes 
for the space and time-like regions indeed differ enough to be measurable. 
The analysis was presented for  the $\rho$ trajectory, where much data are 
available, both in the resonance region and in the space-like region  
through the $\pi N$ charge exchange reaction. 


\end{document}